\documentclass[%
aps,prl
amsmath,amssymb,
reprint,
]{revtex4-2}

\usepackage{float}
\usepackage{graphicx}
\usepackage{dcolumn}
\usepackage{bm}
\usepackage{xcolor}
\usepackage{epstopdf, epsfig, wasysym}
\usepackage{setspace}
\definecolor{mypink1}{rgb}{0.95, 0.95, 0.08}
\usepackage{soul}
\sethlcolor{mypink1}

\begin{document}
	
	\preprint{AIP/123-QED}
	
	\title[Sample title]{The Activation Energy for Wall Slip }
	
	\author{P. F. Pelz}
	\email{peter.pelz@fst.tu-darmstadt.de}
	\affiliation{Technische Universit\"at Darmstadt}
	\author{T. Corneli}%
	\affiliation{Technische Universit\"at Darmstadt}
	
	\date{May 14, 2018}
	
	\begin{abstract}
    The Navier slip boundary condition is interpreted as an equilibrium of shear rate and slip rate. From the argument that the slip rate shall be proportional to the molecules' collision rate, the temperature dependence of the Navier slip boundary condition is derived. The model for the temperature dependence of the slip length is validated by slip measurements of liquid hydrocarbons in a novel Couette typ tribometer being introduced. The essence of the gained experimental data for one fluid-solid-interface is the quadruple activation energy for shear and wall slip together with the viscosity and slip length at a reference temperature. {This quadruple is determined for four different hydrocarbon liquids of different molecular mass, structure and polarity proving the applicability of the new measurement method. From the executed systematic measurements three conclusions regarding the slip length dependence are pointed out: (i) the slip length increases with increasing molar mass; (ii) changing the molecular structure from saturated hydrocarbon to unsaturated affects the slip length as well as the activation energy for slip; (iii) adding a small fraction of polar molecules to the hydrocarbon decreases the slip length and increases the activation energy for wall slip due to the polar end-groups of the liquid.}\\
		%
	\end{abstract}
	
	\pacs{Valid PACS appear here}
	\maketitle

%
%
During the past 20 years, different research groups measured slip at the interface of a fluid moving parallel to a solid; cf. review articles on boundary slip \cite{Neto2005,Lauga2007,Cao2009}.
 Understanding the molecular physics behind the slip condition is of importance for many fields of science: the transfer of momentum and matter as well as heterogeneous reactions in biological or technical systems are greatly influenced by the interface condition which is firstly formulated by Navier \cite{Navier1822} in 1822.\\  
 Assessing the experimental investigations, it strikes that the temperature dependence of the Navier slip condition has been left out of the focus of most research on wall slip \cite{Meinhart1999,Tretheway2002,Joseph2005direct,Joseph2006slippage,Leger1997,Leger2000,Leger2003friction,Pit1999slippage,Pit2000,Schmatko2005,Watnabe1999drag,Cheng2002fluid,Zhu2001,Zhu2002a,Zhu2002c,Zhu2002limits,bonaccurso2003surface,bonaccurso2002hydrodynamic,craig2001shear,Ecke2001measuring,Vinogradova2003dynamic,Restango2002surface,Cottin-Bizonne2002,Cottin-Bizonne2005,Cottin2008Nanohydrodynamics,Neto2003evidence,henry2004effect,Neto2003evidence,henry2004effect,cho2004dipole,Trusdell2006drag,Cheikh2003}. This is unexpected since the temperature dependence of a continuum mechanical property, e.g. the viscosity, gives an insight into the molecular constitution of matter. A prominent example for this is Einstein's work on Brownian motion \cite{Einstein1905}. From thermodynamics, kinetic theory or statistical mechanics the same rate equation 
\begin{eqnarray}
\label{eq_111}
\frac{\mathrm{d}\, \ln k }{\mathrm{d}T}=\frac{E}{\mathcal{R}T^2} + \mathrm{const}.
\end{eqnarray}
is derived, relating the rate $k$ of a molecular process to the temperature $T$, activation energy $E$ and gas constant $\mathcal{R}$ \cite{Laidler1983the}. With the temperature independent activation energy, the well known Arrhenius relation follows $k\propto\exp(-E/\mathcal{R}T)$ for the rate of the molecular process.\\
Before applying and validating the Arrhenius relation for wall slip, and consequently determining the activation energy, {we briefly review the experimental research on slip length measurements. This motivates the new Couette type slip length measuring device introduced here, being develop during the last five years, as an device where the surface averaged slip length is derived from an integral measure, i.e. the torque.}\\
So far, the experimental research focuses either on verifying the existence of a wall slip or on influencing the effect by surfactants. Systematic investigations concentrate on the variation of shear rates. The suspected dependence of slip length on wall shear rate is still under discussion \cite{Zhu2001,Zhu2002a,Zhu2002c,Zhu2002limits,Choi2003apparent,bonaccurso2003surface,craig2001shear}. This might be due to mixing up measurement uncertainty and interpretation: the here presented measurements reveal no shear rate dependence of the slip length for hydrocarbon liquids moving relative to a metal surface up to a shear rate of the order of magnitude $10^5\, \mathrm{s}^{-1}$  in the temperature interval ${12.5\, ^\circ\mathrm{C} \ \mathrm{to} \ 60.2\, ^\circ\mathrm{C}}$. \\ 
{The following liquid-solid-interface combinations are in the focus of research so far: water} \cite{Joseph2005direct,Joseph2006slippage,Watnabe1999drag,Cheng2002fluid,Choi2003apparent,Zhu2001,Zhu2002a,Zhu2002limits,Cottin-Bizonne2002,Cottin-Bizonne2005,Cottin2008Nanohydrodynamics,Lumma2003,Meinhart1999,Tretheway2002,Watnabe1999drag}, alkane \cite{Cheng2002fluid,Zhu2001,Zhu2002a,Zhu2002c,Zhu2001,Cottin-Bizonne2005,Cottin2008Nanohydrodynamics,Pit2000,Pit1999slippage,Schmatko2005} or polymer melts \cite{,Leger1997,Leger2003friction}  {moving relative to glass, mica, silicon} \cite{Joseph2005direct,Joseph2006slippage,craig2001shear,bonaccurso2002hydrodynamic,bonaccurso2003surface,Vinogradova2003dynamic,Zhu2001,Zhu2002a,Zhu2002c,Zhu2002limits,Restango2002surface,Cottin-Bizonne2002,Cottin-Bizonne2005,Cottin2008Nanohydrodynamics,Meinhart1999,Tretheway2002,Huang2006direct,Leger1997} {or synthetic sapphire}  \cite{Pit1999slippage,Pit2000,Leger2003friction,Schmatko2005}. {Technical most relevant is the hydrocarbon-metal-interface beeing here in focus. Keeping in mind, that roughly one third of primary energy is dissipated due to friction at this liquid-solid-interface} \cite{Spurk1992,Holmberg2012,Holmberg2014} {the relevance of the hydrocarbon-metal-interface becomes clear.}\\
{Until now, a systematic investigation of the temperature influence on the slip length aiming at the Arrhenius relation has no bee pursued. Most of the slip length data are reported for only one temperature. Only Churaev et. al.} \cite{Churaev1984slippage} {and Fetzer et al.} \cite{Fetzer2006,Fetzer2007} {measure slip lengths at different temperatures. First steps are made in simulating the molecular dynamics at the interface} \cite{Mueller2008flow,Servantie_2008}: {although the dimensionless temperature is varied, the experimental validation of the simulation is still pending, since length and energy scales of the molecules are so far unknown. In this light, the presented results may contribute in determining some of the unknown scales for future molecular dynamic simulations. Churaev et al.} \cite{Churaev1984slippage} {measure the slip length of water moving relative to solid glass in the temperature range from ${2\, ^\circ\mathrm{C}}\ \mathrm{to}\ 35\, ^\circ\mathrm{C}$. Fetzer et al.} \cite{Fetzer2006,Fetzer2007} {measure the slip length of an amorphous liquid polymer melt moving relative to a coated silicon solid. The temperature ranges from ${105\, ^\circ\mathrm{C} \ \mathrm{to} \ 130\, ^\circ\mathrm{C}}$, i.e. close to the glass transition temperature of the polymer. Both measurements reveal a decreasing slip length with increasing temperature. Fetzer et al. fit a power law for the temperature dependent friction coefficient without gaining further physical insight.}\\
{Within this letter we address five novel findings to the physics at the interface of a liquid and a solid wall: (i) by arguments based on dimensional analysis and the collision model we derive an Arrhenius relation for the temperature dependence of the slip length; (ii) we present a novel Couette type tribometer being developed over the last five years to measure wall slip at various temperatures; (iii) the presented experiments reveal no shear rate dependence of the slip length up to a shear rate of the order of magnitude $10^5\, \mathrm{s}^{-1}$  in the temperature interval ${12.5\, ^\circ\mathrm{C} \ \mathrm{to} \ 60.2\, ^\circ\mathrm{C}}$; (iv) the Arrhenius relation for the temperature dependence of the slip length is experimental validated for four hydrocarbon liquids of different molecular masses and the activation energy for wall slip is derived for all four interfaces; (v) the influence of molecular weight, structure and polarity of the fluid on the slip length and activation energy is discussed.}\\
\begin{figure}
	\begin{minipage}[t]{0.4\textwidth}
		\def\svgwidth{60mm}
		\centerline{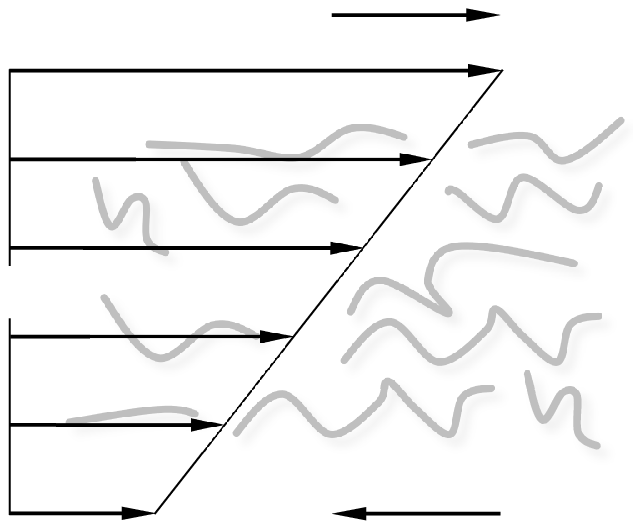}
	\end{minipage}\\
	\vspace{0cm}
	\vspace{-1cm}
	\begin{minipage}[t]{0.4\textwidth}  
		\def\svgwidth{60mm}
		\centerline{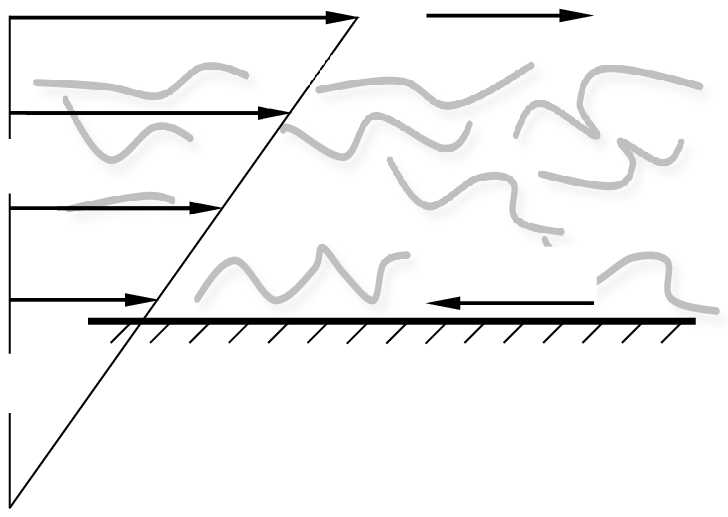}
		\caption{Schemata of bulk shear (a) and wall slip (b) for molecules sliding relative to each other and relative to a solid wall.}
		\label{figure_1}
	\end{minipage}
\end{figure}
In the following paragraph we motivate the Arrhenius relation for wall slip from collision theory. Shear and slip at a solid wall are related by the Navier slip boundary condition: although this boundary condition is often seen as a kinematic relation, Navier himself has interpreted it in the year 1822 as a dynamic relation \cite{Navier1822}. As will be seen, this original view indeed leads to deeper insight. 
Figure \ref{figure_1} shows schematically (a) bulk shear deformation and (b)  wall slip of molecules at homogeneous temperature $T$. At equilibrium the near wall molecules adhere to the wall, $y=0$, as well as to each other. In the ensemble average the electrostatic molecular forces sum up to a shear stress $\tau$ being a macroscopic or continuum mechanical quantity such as dynamic viscosity $\mu(T)$. At non-equilibrium the molecules slide (in the time average) relative to each other and relative to the wall. $u(y)$ denotes the ensemble averaged velocity parallel to the wall at the wall normal distance $y$.\\ 
Applying a shear stress $\tau$ to a Newtonian fluid at constant temperature $T$ results in a shear deformation of constant rate $\partial u/\partial y=\tau/\mu(T)$. The molecular counterpart to this ensemble averaged macroscopic rate is the collision rate of the molecules $k_\mu(T)\propto \exp (-E_\mu/\mathcal{R}\,T)$ (activation energy $E_\mu$, general gas constant $\mathcal{R}$). For reasons of dimensions, i.e. due to the Bridgman’s postulate \cite{Bridgman1922dimensional}, both rates shall be linear dependent resulting in $\tau/\mu(T)\propto \exp (-E_\mu/\mathcal{R}\,T)$ (cf. the argumentation of Truesdell \cite{Truesdell1952}). Hence, the well known temperature dependence of viscosity of a liquid $\mu(T)\propto \exp (E_\mu/\mathcal{R}\,T)$ is on hand. The logarithm of the ratio of $\mu(T)$ and $\mu_0=\mu(T_0)$ defines the time-temperature-shift factor
\begin{equation}
\label{eq_3}
a_\mu: = \log{\frac{\mu(T)}{\mu(T_0)}}=\frac{E_\mu}{\mathcal{R}}\bigg(\frac{1}{T}\!-\!\frac{1}{T_0}\bigg).
\end{equation}
With this in mind, the expectation and motivation for the research presented here is the following: there should be an Arrhenius relation  such as (\ref{eq_3}) and hence an activation energy $E_\lambda$ for wall slip as well. Based on Navier's work, Helmholtz \cite{Helmholtz1860} introduced in 1860 the slip length ${\lambda}(T)$ as the ratio of slip velocity and velocity gradient $u(0) /(\partial u/\partial y)\big|_{y=0}$. {Thus, besides the shear rate} $\tau/\mu(T)$, {there is a second ensemble averaged macroscopic rate, the slip rate}  $u(0)/{\lambda}(T)$. {The molecular counterpart to this rate is the collision rate of molecules at the interface}  $k_\lambda \propto \exp (-{E}_\lambda/\mathcal{R}\,T)$, {being different from} $k_\mu$. {The slip rate and} $k_\lambda$ {are linear dependent due to reasons of dimensions resulting in} $u(0)/\lambda(T)\propto \exp (-E_\lambda/\mathcal{R}\,T)$. {Hence, we derive the temperature dependence of the slip length of a liquid} $\lambda(T)\propto \exp (E_\lambda/\mathcal{R}\,T)$. {It is expected, that the activation energy for wall slip} $E_\lambda$ {differs from the activation energy} $E_\mu$ {for bulk shear. The logarithm of the ratio of}  $\lambda(T)$ and $\lambda_0=\lambda(T_0)$ {defines again a time-temperature-shift factor}
\begin{equation}\label{eq_4}
a_\lambda := \log{\frac{\lambda(T)}{\lambda(T_0)}}=\frac{E_\lambda}{\mathcal{R}}\bigg(\frac{1}{T}\!-\!\frac{1}{T_0}\bigg),
\end{equation}
{but now for wall slip. In conclusion, the viscous momentum transport for one fluid-solid-interface is fully determined by the provided quadruple $[\mu(T_0),\, E_\mu,\, \lambda(T_0),\, E_\lambda]$. Equivalent to $\lambda(T)$, sometimes the ratio $k(T):=\lambda(T)/\mu(T)=u(0) / \tau$ is used cf.} \cite{Fetzer2006,Servantie_2008,Mueller2008flow}. {This so called friction factor of course follows an Arrhenius relation as well, showing the activation energy  $E_k=E_\lambda-E_\mu$. The quadruple $[\mu(T_0),\, E_\mu,\, \lambda(T_0 ),\, E_\lambda]$ and $[\mu(T_0),\, E_\mu,\, k(T_0),\, E_k]$ are equivalent; with one given quadruple the other one can be derived by the given transformation.\\}
In the light of the above discussion, it is indeed beneficial interpreting the Navier slip boundary condition as a dynamic equilibrium of the two rates, i.e. slip rate and shear rate:
\begin{equation}\label{eq_1a}
\frac{u(0)}{\lambda(T)}=\frac{\tau(0)}{\mu(T)}\ \mathrm{(at\ the\ wall).}  
\end{equation}
With (\ref{eq_3}) and (\ref{eq_4}) the temperature dependence of this dynamic Navier slip boundary condition is expected to follow the relation
\begin{equation}\label{eq_Rates}
\log\frac{\tau(0)}{\mu_0}\frac{\lambda_0}{u(0)}=\frac{a_\mu(T,T_0)}{a_\lambda(T,T_0)}=\frac{E_\mu\!-\!E_\lambda}{\mathcal{R}}\bigg(\frac{1}{T}-\frac{1}{T_0}\bigg). 
\end{equation}
%
%
%
%
%
%
%
%
%
%
To validate the Arrhenius relation for wall slip, we introduce a new integral method for measuring the bulk viscosity $\mu(T)$ and wall slip $\lambda(T)$ as a function of the fluid temperature $T$ at once.
%
\begin{figure}
    \def\svgwidth{70mm}
	\centerline{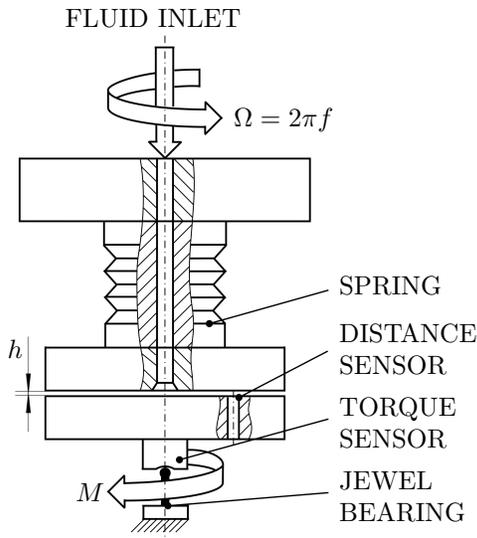}
	\caption{\label{fig:epsart} Principle sketch of the slip length tribometer (disk diameter is $64\ \mathrm{mm}$). The temperature is controlled by tempering the tribometer as well as the fluid in a temperature test chamber.}
	\label{fig_1}
\end{figure}
The apparatus schematically shown in figure \ref{fig_1} is the slip length tribometer developed over the last five years by the authors: the upper disk rotates with constant rotational speed $\Omega=2\pi f$ relative to the lower one. Both circular disks $(\diameter 64\ \mathrm{mm})$ are made of an edge hardened stainless steel (steel type 1.8519).The planarity of both disks is smaller than $\pm15\ \mathrm{nm}$ and the arithmetic mean roughness is below $10\ \mathrm{nm}$. The transmitted torque $M$ is measured at the support of the lower disk. The gap height $h$ is measured by means of a capacitive distance sensors with a resolution of $4\ \mathrm{nm}$.\\ 
To allow a cardanic self levelling of the two disks relative to each other, the lower one is supported by a jewel bearing. The gap height $h$ is controlled by means of the fluid pressure at the fluid inlet. The pressure forces the fluid radial outward. At the same time the fluid is sheared in the circumferential direction. The circumferential Couette velocity profile is $u(r,y)=\Omega r\,  (y\!+\!\lambda)/(h\!+\!2\lambda)$ (radius $r$, axial coordinate $0\!<\!y\!<\!h$). For small gap height $h$ the nonlinear convective acceleration is negligible and the equation of motion is linear. Hence, the circumferential Couette velocity profile is independent from the radial velocity profile. Thus, for the laminar flow the inverse torque is gained as 
\begin{eqnarray}\label{eq_5}
M^{-1}=\frac{h\!+\!2\lambda}{\mu\,\Omega I_P},
\end{eqnarray}
with the polar second moment of area $I_P:=\int_A r^2\ \mathrm{d}A$ of the disks.\\ 
The relation (\ref{eq_5}) shows that the inverse torque $M^{-1}$ is proportional to the apparent gap height $h\!+\!2\lambda$. $M^{-1}$ is linear in $h$.  For the no-slip boundary condition, i.e. vanishing slip length $\lambda\to 0$, the inverse torque $M^{-1}\to 0$ for $h\to 0$. For non-vanishing slip length $\lambda>0$ the $M^{-1}-h-\mathrm{curve}$ is shifted to the left. The intersection with the axis of abscissa equals two times the slip length $\lambda$ (cf. fig. \ref{results_PAO6_40}). It is not possible to reach $h\to0$ due to two reasons: first, due to micro roughness a solid-solid contact would not be avoidable; second, $M$ would tend to infinity as $h$ approaches zero. But this is no drawback: the conceptional design of the device is such, that $\lambda$ is gained from the extrapolation of the line relation $h\to 0$.
%
%
%
\begin{figure}
		\def\svgwidth{65mm}
		\centerline{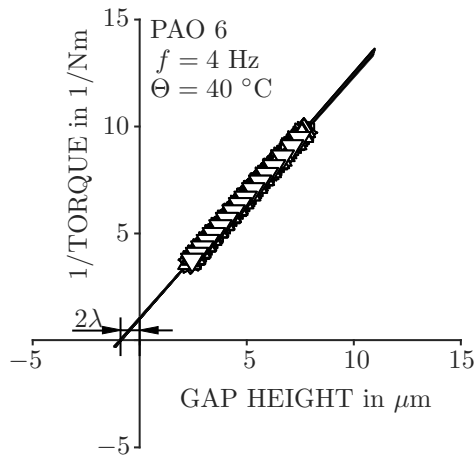}
		\caption{Result of a typical experiment at $\Theta=40 \ ^\circ$C and constant rotational speed of $4\ \mathrm{Hz}$ for a synthetic hydrocarbon alpha-olefin (PAO 6). The figure shows 42 measurement runs. Each run is marked by a symbol. Only the last one marked by triangles is visible.}		
		\label{results_PAO6_40}
\end{figure}
%
%
%
%
%
Figure \ref{results_PAO6_40} shows the result of typical experiments at constant temperature of $T=(313.15 \pm 0.1)\ \mathrm{K}$, i.e. at $\Theta=(40\pm0.1)\ ^\circ\mathrm{C}$. Clearly the linear behaviour of the inverse frictional torque $M^{-1}$ reveals. The coefficient of determination for equation (\ref{eq_5}) is above $0.999$. 
{Based on this linearity of the inverse torque, the earlier mentioned shear rate independence of the slip length is asserted: all measurements plotted in Fig.} \ref{results_PAO6_40} {show the same slip length, even though the shear rate varies from} ${2\cdot10^3\,\mathrm{s}^{-1}\ \mathrm{to}\ 2\cdot10^5\,\mathrm{s}^{-1}}$. {The measured mean slip length for synthetic hydrocarbon alpha-olefin (PAO 6) at} $40\,^\circ\mathrm{C} $ is $\lambda =475\ \mathrm{nm}$ {(for a detailed uncertainty quantification the reader is referred to the supplementary material of this letter).}\\
%
%
%
Figure \ref{fig_ArrheniusPlot} shows the Arrhenius plot for the time-temperature-shift factors (\ref{eq_3}), (\ref{eq_4}) for bulk shear $a_\mu(T,T_0)$ and wall slip $a_\lambda(T,T_0)$: both factors are plotted versus the inverse absolute temperature $T$. As expected, both relations (\ref{eq_3}) and (\ref{eq_4}) are observed validating the model. The energy barrier for bulk shear of the alpha-olefin is $E_\mu=33.5\ k\mathrm{J}\, \mathrm{mol}^{-1}$.  New is the so far unknown activation energy for slip relative to lipophilic metal surface: the experiments indicate an activation energy of $E_\lambda=17.6\ k\mathrm{J}\, \mathrm{mol}^{-1}$.
\begin{figure}
	\def\svgwidth{95mm}
	\centerline{\input{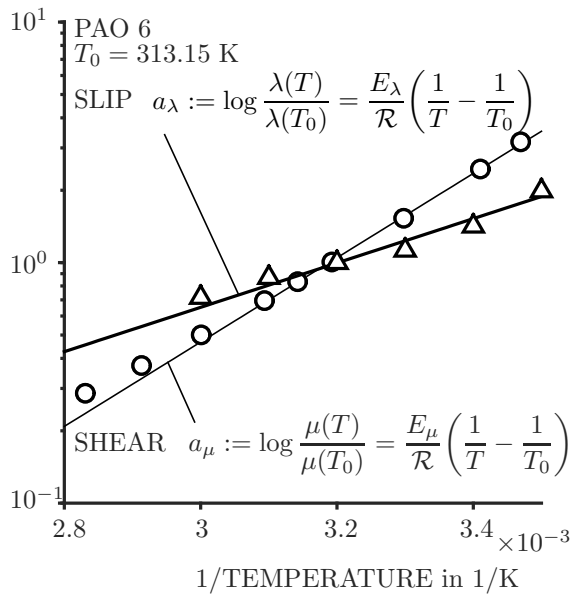}}
	\caption{\label{fig_ArrheniusPlot} Arrhenius plot for bulk shear and wall slip for the alpha-olefin. The activation energy for bulk shear is $E\mu=33.5\ k\mathrm{J}\, \mathrm{mol}^{-1}$ and for wall slip $E_\lambda=17.6\ k\mathrm{J}\, \mathrm{mol}^{-1}$.}
\end{figure}\\
{The activation energy for bulk shear and wall slip shall both depend on the van der Waals forces and hence on two major factors: first on the molecule's length, i.e. the geometric mean average of the molar mass $\overline{M}$; second on the polarity of the molecules.}\\
{To validate this hypotheses, we varied the molecular structure from the unsaturated hydrocarbon alpha-olefin, being a blend of three main fractions} (cf. Fig. \ref{fig_MolarMass}), {to two saturated hydrocarbons, i.e. mineral oils of mean molar mass $\overline{M}=666\, \mathrm{u}$ (ISO VG 46) and $806\, \mathrm{u}$ (ISO VG 68). Finally we added a small fraction of long chained molecules with polar end-groups (viscosity index improver, VI) to the mineral oil (ISO VG 46). Adding a small fraction of long chained molecules changes the viscosity as well as the activation energy for bulk shearing. The fraction of the added molecules was such selected that the viscosity of the modified mineral oil (ISO VG 46+VI) at $40\,^\circ \mathrm{C}$ corresponds to the viscosity of the mineral oil ISO VG 68 at the same temperature.}\\
Table \ref{tab_results} {gives the quadruple $[\mu(T_0),\, E_\mu,\, \lambda(T_0 ),\, E_\lambda]$ , describing the tribological system hydrocarbon-metal-interface for each fluid. From the results we point out three conclusions regarding the dependence of the slip length and activation energy for slip: (i) the slip length increases with increasing molar mass; (ii) a change from saturated (eq. ISO VG 68) to unsaturated (eq. PAO 6) hydrocarbon influences both shear and slip; (iii) adding a small fraction of polar molecules to the hydrocarbon decreases the slip length and increases the activation energy for wall slip due to the polar end-groups of the liquid. Assessing the unpolar hydrocarbon molecules, there is an increase in slip length with increasing molecular mass.}\\
\\
\begin{table}[h!]
	\caption{\label{tab_results}%
		The table gives the quadruple $[\mu(T_0),\, E_\mu,\, \lambda(T_0 ),\, E_\lambda]$ for each of the four investigated fluids. }
	\begin{ruledtabular}
		\begin{tabular}{llllll}
			\textrm{Fluid}&
			\textrm{$\overline{M}$}&
			\textrm{$\mu(T_0)$}&
			\textrm{$E_\mu$}&
			\textrm{$\lambda(T_0)$}&
			\textrm{$E_\lambda$}\\
			\textrm{}&
			\textrm{in}&
			\textrm{in}&
			\textrm{in}&
			\textrm{in}&
			\textrm{in}\\
			\textrm{}&
			\textrm{u}&
			\textrm{mPas}&
			\textrm{kJ\,mol$^{-1}$}&
			\textrm{nm}&
			\textrm{kJ\,mol$^{-1}$}\\

			\colrule
			ISO VG 46      & 666 & 38.4  & 41.3  & 120  & 23.7\\
			ISO VG 68      & 806 & 58.1  & 45.0  & 143  & 26.7\\
			\colrule
			PAO 6          & 882 & 25.8  & 33.5  & 468  & 17.6\\
			\colrule
			ISO VG 46 + VI & 666 & 58.1  & 40.3  & 65   & 43.0\\
		\end{tabular}
	\end{ruledtabular}
\end{table}\\
%
%
%
\begin{figure}[h!]
	\vspace{10mm}
	\def\svgwidth{60mm}
	\centerline{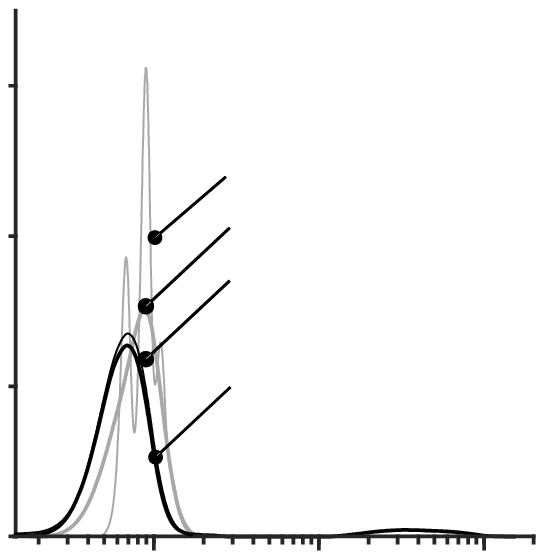}
	\caption{\label{fig_MolarMass} Area normalized molar mass distribution for the alpha-olefin (PAO 6), the mineral oils (ISO VG 46, ISO VG 68) as well as for the mineral oil modified by a small fraction of long
		chained molecules with polar end-groups (ISO VG 46 + VI).}
\end{figure}

	\nocite{*}
	\bibliography{Literature_PhysicalReviewLetters}

	\clearpage

	\section*{Supplementary material for "The Activation Energy for Wall Slip": Uncertainty Quantification}

\begin{figure}[h!]
	\def\svgwidth{60mm}
	\centerline{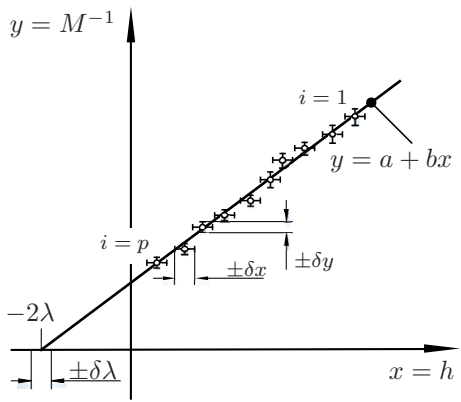}
	\caption{\label{fig:uncertainty} Statistical and systematic uncertainties for measuring the slip length.}
\end{figure}

In this supplementary material for the letter "The Activation Energy for Wall Slip" we provide a rigorous uncertainty quantification for the measurement data gained with the novel Couette type tribometer being introduced. The presented uncertainty quantification leads to the confidence interval for the slip length $\lambda$, i.e. $\delta\lambda_{95\%}=\pm 91.4\,\mathrm{nm}$ and the uncertainty of the measurement system $\delta\lambda=\pm 33.2\,\mathrm{nm}$.\\
The slip length tribometer is based on the linear relation (cf. Eq. (6) of the letter)

\begin{eqnarray}\label{eq:gerade}
M^{-1}=\frac{h\!+\!2\lambda}{\mu\Omega I_P}\qquad \mathrm{or} \qquad y=a+bx,
\end{eqnarray}
mapping the distance $x=h$ on the inverse torque $y=M^{-1}$. In the experiment, the gap height $h$ and the frictional torque $M$ are the directly and simultaneously measured quantities. The indirect measured quantities are the sum of the slip length $2\lambda$ at both interfaces, given by the zero point of the linear equation (cf. Fig. 3 of the letter) and the viscosity $\mu$. Provided $a$ and $b$ are unknown, the slip length is obtained by 
\begin{eqnarray}\label{eq:slip_length}
\lambda = \frac{1}{2}\frac{a}{b}.
\end{eqnarray}
The viscosity is 
\begin{eqnarray}\label{eq:slip_viscosity}
\mu = \frac{1}{b\,\Omega I_P}.
\end{eqnarray}
The intercept 
\begin{eqnarray}
\label{eq:intercept}
a = \frac{\sum x_i^2 \sum y_i -\sum x_i \sum x_i y_i}{n\sum x_i^2-\big(\sum x_i\big)^2},\quad  i=1,\dots,n,
\end{eqnarray}
and the slope
\begin{eqnarray}
\label{eq:slope}
b = \frac{n\sum x_i y_i -\sum x_i\sum y_i}{n\sum x_i^2-\big(\sum x_i\big)^2},\quad  i=1,\dots,n,
\end{eqnarray}
of Eq. (\ref{eq:gerade}) are determined by means of a least square fit for the $n$ measured data. Since both measurands, the distance $x$ and the inverse torque $y$, are subjected to a systematic and statistical measurement uncertainty, intercept $a$ and the slope $b$ are uncertain as well. The square of the uncertainty regarding the slip length (cf. Eq. (\ref{eq:slip_length})) is given by the Gaussian uncertainty propagation
\begin{eqnarray}
\label{eq:uncertainty_slip_1}
(\delta\lambda)^2 &=&\bigg(\frac{\partial \lambda}{\partial a}\bigg)^2(\delta a)^2 +\bigg(\frac{\partial \lambda}{\partial b}\bigg)^2(\delta b)^2\\
\label{eq:uncertainty_slip_2}
&=& \bigg(\frac{1}{2b}\bigg)^2(\delta a)^2 +\bigg(\frac{a}{2 b^2}\bigg)^2(\delta b)^2.
\end{eqnarray}
Eq. (\ref{eq:uncertainty_slip_2}) considers the uncertainty of the intercept $\delta a$ as well as the uncertainty of the slope $\delta b$. Both uncertainties are unknown so far. The uncertainties of intercept and slope are determined using the generalized Gaussian uncertainty quantification. Both direct measurands show a mean $\overline{x}$, $\overline{y}$, a statistical uncertainty $s$ (empirical standard deviation) and systematic uncertainty $\delta x$, $\delta y$ (cf. Fig. \ref{fig:uncertainty}).  In the experiment $x$ and $y$ are independent measurands. Thus, with respect to correlated uncertainty, only the systematic measurement uncertainties have to be taken into account as correlated uncertainties. The uncertainty of intercept $\delta a$ and slope $\delta b$ are than given by the following expressions:
\begin{eqnarray}\label{eq:uncetainty_a}\hspace{-0.5cm}
 (\delta a)^2&=&\sum_{i=1}^{n}\Bigg[(\delta x)^2 \bigg(\frac{\partial a}{\partial x_i}\bigg)^2+\Big[s^2\!+\!(\delta y)^2\Big]\bigg(\frac{\partial a}{\partial y_i}\bigg)^2\Bigg]+\\ \nonumber &+&
2\sum_{i=1}^{n-1}\sum_{k=i+1}^{n}\Bigg[(\delta x)^2 \bigg(\frac{\partial a}{\partial x_i}\bigg)\bigg(\frac{\partial a}{\partial x_k}\bigg) + \\ \nonumber &+& (\delta y)^2 \bigg(\frac{\partial a}{\partial y_i}\bigg)\bigg(\frac{\partial a}{\partial y_k}\bigg)\Bigg],\\
\label{eq:uncetainty_b}\hspace{-0.5cm}
(\delta b)^2 &=&\sum_{i=1}^{n}\Bigg[(\delta x)^2 \bigg(\frac{\partial b}{\partial x_i}\bigg)^2+\Big[s^2\!+\!(\delta y)^2\Big]\bigg(\frac{\partial b}{\partial y_i}\bigg)^2\Bigg]+\\ \nonumber &+&
2\sum_{i=1}^{n-1}\sum_{k=i+1}^{n}\Bigg[(\delta x)^2 \bigg(\frac{\partial b}{\partial x_i}\bigg)\bigg(\frac{\partial b}{\partial x_k}\bigg) +\\ \nonumber &+& (\delta y)^2 \bigg(\frac{\partial b}{\partial y_i}\bigg)\bigg(\frac{\partial b}{\partial y_k}\bigg)\Bigg].
\end{eqnarray}
The partial derivatives are given for $i,j=1,\dots,n$ by
\begin{eqnarray}
\label{eq_da_dx}
\frac{\partial a}{\partial x_i} &=&  \frac{2x_i\sum y_j -\big(\sum x_j y_j +y_i\sum x_j \big) }{n\sum x_j^2-\big(\sum x_j\big)^2} + \\ \nonumber  &-& 2\frac{\Big(\sum x_j^2 \sum y_j -\sum x_j \sum x_j y_j\Big)\Big(n x_i-\sum x_j\Big)}{\Big(n\sum x_j^2-\big(\sum x_j\big)^2\Big)^2},\\ 
\label{eq_da_dy}
\frac{\partial a}{\partial y_i} &=& \frac{\sum x_j^2-x_i\sum x_j}{n\sum x_j^2-\big(\sum x_j\big)^2},\\ 
\label{eq_db_dx}
\frac{\partial b}{\partial x_i} &=&  \frac{n\, y_i -\sum y_j }{n\sum x_j^2-\big(\sum x_j\big)^2}+ \\ \nonumber  &-&2\frac{\Big(n\sum x_j y_j -\sum x_j\sum y_j \Big)\Big(n x_i-\sum x_j \Big)}{\Big(n\sum x_j^2-\big(\sum x_j\big)^2\Big)^2},\\
\label{eq_db_dy}
\frac{\partial b}{\partial y_i} &=& \frac{n\,x_i -\sum x_j}{n\sum x_j^2-\big(\sum x_j\big)^2}.
\end{eqnarray}
Still the statistical uncertainty $s$ (empirical standard deviation) and systematic uncertainties $\delta x$ and $\delta y$ have to be considered. The origin of these uncertainties as well as their quantification are explained step by step in the following section.\\
\\
The empirical variance 
\begin{eqnarray}
\label{eq:emp_var}
s^2=\frac{1}{n\!-\!1}\sum_{i=1}^{n}\Big[M_i^{-1}-(a+bx_i)\Big]^2
\end{eqnarray}
takes the statistical uncertainty of the measurands $(x_i,y_i)$ from the linear relation $y=a+bx$ into account. This statistical uncertainty is reduced by repeating the experiment $m$ times. We perform $m= 20 \dots 40$ measurement series to reduce $(\delta \lambda)^2$. Each measurement series consists of $p=15\dots 20$ measuring points, cf. Fig. \ref{fig:uncertainty}. The evaluation of the linear regression is based on the measuring points of all $m$ measurement series. Hence, the linear regression is supported by $n=m*p=300\dots800$ measuring points for one single slip length measurement only. This results in a variance $s^2\approx 0.0015\, (\mathrm{Nm})^{-2}$ for Eq. (\ref{eq:emp_var}). In addition, each measuring point is averaged over $70,000$ individual measurements in a measurement interval of 10 seconds. As a results of this averaging process, the variances of the single measuring points can be neglected compared to the deviations of the linear relation. The measurement time per measurement series to obtain the slip length is approximately 45 minutes. On average, 30 series of measurements are recorded for each temperature dependent slip length in the Arrhenius plot (cf. Fig. 4 of the letter).\\  
The systematic uncertainty $\delta y$ of the inverse torque results from the nonlinearity of the torque measuring system. Since we measure the torque and not the inverse torque, the uncertainty of the inverse torque is determined again by means of Gaussian uncertainty propagation
\begin{eqnarray}
\label{eq:delta_y_M_inv_errorbound}
\delta y=\frac{1}{M^2}\delta M.
\end{eqnarray}
The systematic uncertainty of the torque measurement is composed of the uncertainty of the sensor and the signal processor and is specified by an uncertainty bound $\delta M=0.7\,\mathrm{mNm}$.  An uncertainty bound corresponds to a uniform distribution of the uncertainty. For the Gaussian uncertainty propagation, the uncertainty shall be normal distributed. Hence, the uniform distributed uncertainty of the uncertainty bound is converted into a normal distributed uncertainty 
\begin{eqnarray}
\label{eq:delta_y_M_inv}
\delta y=\frac{1}{\sqrt{3}}\frac{\delta M}{M^2},
\end{eqnarray}
cf. Kamke \cite{Kamke2010}.\\
The systematic uncertainty $\delta x$ of the distance measurement results from two sources: (i) the calibration of the capacitive distance sensor $\delta x_\mathrm{S}=60\,\mathrm{nm}$, and (ii) the planarity of the disks $\delta x_\mathrm{D}=30\,\mathrm{nm}$. The total uncertainty of the distance measurement results from the addition of both partial uncertainties $\delta x=90\,\mathrm{nm}$.\\
We are now in a position for determining the total measurement uncertainty of the slip length measurement in two steps: (i) first  the statistical, $s$, and the systematic measurement uncertainties, $\delta x$, $\delta y$, are inserted into equations (\ref{eq:uncetainty_a}) and (\ref{eq:uncetainty_b}), determining the uncertainty of intercept $\delta a$ and slope $\delta b$; (ii) second, the calculated uncertainty of intercept $\delta a$ and slope $\delta b$ are inserted in Eq. (\ref{eq:uncertainty_slip_2}) giving the uncertainty of the slip length measurement $\delta \lambda$.\\ For the measurement presented in Fig. 3 of the letter the uncertainty amounts for $\delta\lambda=46.6\,\mathrm{nm}$. Considering a $95\%$ confidence interval the slip length is measured with an uncertainty of $\delta \lambda_{95\%}=\pm91.4\,\mathrm{nm}$.\\
Reported uncertainties for slip length measurements vary of the order of magnitude $1\, \mathrm{nm}$ to $10^3\, \mathrm{nm}$. Cottin-Bizonne et al. \cite{Cottin-Bizonne2005} measure the slip length with an uncertainty below  $10\, \mathrm{nm}$ for atomically smooth surfaces. The reported uncertainty by Pit et al. \cite{Pit2000} is  $50\,\mathrm{nm}$ for their experiments. The reported uncertainty by Fetzer et al. \cite{Fetzer2007,Fetzer2006} varies between $50\dots 3000\,\mathrm{nm}$.\\ 
Our uncertainty consideration takes the uncertainty of the probe as well as the uncertainty of the measurement system into account, since the uncertainty  $\delta x_\mathrm{D}=30\,\mathrm{nm}$ is related to the manufacturing tolerance of the metal surfaces. Excluding this uncertainty, the uncertainty of the measurement system yields $\delta \lambda=33.2\,\mathrm{nm}$. This uncertainty is relevant for comparing the tribometer as a measurement system with measurement systems measuring the slip length of atomically smooth surfaces.\\ 
In conclusion, the introduced novel Couette typ tribometer is indeed reliable and competitive for measuring the temperature-dependent slip length of atomically smooth as well as of rough surfaces.

	\begin{acknowledgments}
		The authors thank Johannes Emmert and Professor Andreas Dreizler, both Technische Universit\"at Darmstadt for the fruitful discussion leading to this uncertainty quantification. 
	\end{acknowledgments}

\end{document}